# Masses and Decay Constants of Heavy-Light Mesons Using the Multistate Smearing Technique*†

Anthony Duncan,[a] Estia Eichten,[b] Jonathan M. Flynn,[c] Brian R. Hill,[d] and Hank Thacker.[e]

[a]Dept. of Physics, Univ. of Pittsburgh, Pittsburgh, PA 15260 USA

[b]Fermilab, MS 106, PO Box 500, Batavia, IL 60510 USA

[c]Physics Dept., Univ. of Southampton, Southampton SO9 5NH UK

[d]Dept. of Physics, Univ. of California, Los Angeles, CA 90024 USA

[e]Dept. of Physics, Univ. of Virginia, Charlottesville, VA 22901 USA

We present results for $f_B$ and masses of low-lying heavy-light mesons. Calculations were performed in the quenched approximation using multistate smearing functions generated from a spinless relativistic quark model Hamiltonian. Beta values range from 5.7 to 6.3, and light quark masses corresponding to pion masses as low as 300 MeV are computed at each value. We use the $1P$–$1S$ charmonium splitting to set the overall scale.

## 1. INTRODUCTION

Lattice gauge calculations of heavy-light meson structure are of both theoretical and phenomenological interest.[1] One immediate goal of these calculations is to obtain precise quantitative results for masses, decay constants, and form factors in the static approximation, where the heavy quark propagator is replaced by a timelike Wilson line. One difficulty which plagued early, exploratory calculations of the pseudoscalar decay constant $f_B$ was the problem of isolating the ground state contribution to the propagator of the local weak current. Because of the proximity of excited states and their sizable overlap with the local current, a large separation in time was required, with an accompanying loss of statistics. Recent attempts to overcome this problem have employed nonlocal $\bar{Q}q$ operators smeared either over a cube[4] or wall source[5] in a fixed gauge or by gauge invariant methods[6,7]. By measuring the asymptotic behavior of both the smeared-smeared ($SS$) and smeared-local ($SL$) propagators, one can reduce the systematic error associated with excited state contributions.

Here we report results obtained using the multistate smearing method discussed elsewhere[8,9]. The present results offer a wide range of lattice spacings and quark masses from which an extrapolation to the chiral and continuum limits can be obtained.

## 2. RELATIVISTIC QUARK MODEL

The problem of overlap with excited states is reduced by using smearing functions that are carefully chosen to resemble the wavefunctions of lattice QCD. The success of a simple relativistic quark model Hamiltonian in reproducing the lattice wavefunctions greatly simplifies this task. The basic features of this model are:
(a) the use of a relativistic kinetic term $\sqrt{p^2+m^2}$ (with $m$ a constituent quark mass) for the kinetic piece of the Hamiltonian, and
(b) a confining potential $V(\vec{r})$, which is taken to be the static interaction energy obtained from correlators of temporal Wilson lines in lattice QCD.

Since the potential is directly measured on the lattice, the only adjustable parameter in the

*Based on talks presented by E. Eichten, B. Hill, and H. Thacker

RQM is the constituent quark mass $m$. After fixing $m$ by a match to the 1S wavefunction, we have found good agreement with measured excited state wavefunctions (for example, the 1P state). This agreement suggests that this ansatz accurately describes at least the valence quark sector of the full mesonic bound-state.

To minimize lattice discretization and finite volume artifacts in the comparison of RQM and lattice Monte Carlo results, we have generated a set of lattice smearing functions by solving a discretized version of the relativistic Schrodinger equation, on lattices of the same size as those used in the respective Monte Carlos, and in each case with the static potential determined from Wilson line correlators in the same gauge configurations used to extract our quenched QCD results.

## 3. MULTISTATE ANALYSIS

Our object in this section is to outline a general procedure for extracting the maximum usable information from the multistate coupling matrix:

$$C^{ab}(T) \equiv \sum_{\vec{r}\vec{r}'} \Psi^{(a)}_{\rm smear}(\vec{r})\Psi^{(b)}_{\rm smear}(\vec{r}')$$
$$\cdot < 0 \mid \bar{q}(\vec{r},T)Q(0,T)Q^{\dagger}(0,0)q(\vec{r}',0) \mid 0 > \quad (1)$$

where $q(Q)$ are light (heavy) quark operators, and the $\Psi^{(a)}_{\rm smear}$ ($a$=1,2,...$N$) contain the set of orthonormal smearing functions obtained from the RQM as described in the preceding section. From a set of $N_c$ decorrelated gauge configurations, we begin with a correponding ensemble of $N_c$ statistically independent $C^{ab}(T)$ matrices, from which a standard deviation matrix $\sigma^{ab}(T)$ can be obtained directly. In addition to the smearing wavefunctions of the relativistic potential model, the set $\{\Psi^{(a)}\}$ also includes typically the local source generating the desired heavy-light axial-vector matrix element for extracting $f_B$. In the correlator matrix above, the heavy and light quark propagators in each gauge configuration are computed in Coulomb gauge. As we are dealing with global color singlet states on each time slice (color sums are suppressed) $C^{ab}$ is well-defined and non-zero.

Defining states

$$\mid \Phi^a, T > \equiv \sum_{\vec{r}} \Psi^{(a)}_{\rm smear}(\vec{r})Q^{\dagger}(0,T)q(\vec{r},T) \mid 0 > \quad (2)$$

we have

$$C^{ab}(T) = < \Phi^a, T \mid \Phi^b, 0 >$$
$$= \sum_{n=1}^{M} e^{-E_nT} < \Phi^a \mid n >< n \mid \Phi^b > \quad (3)$$
$$+O(e^{-E_{M+1}T})$$

where the states $\mid n >$ are exact eigenstates of the lattice Coulomb gauge transfer matrix with eigenvalues $e^{-E_n}$. The remainder term of order $e^{-E_{M+1}T}$ will of course be small at large Euclidean time, but in addition should have a small prefactor to the extent that our smearing functions $\Psi^{(a)}_{\rm smear}(\vec{r})$ ($a$=1,2,..$M$) do a good job in representing the valence quark structure of the low-lying states, and to the extent that more complicated Fock states (containing extra quark pairs, real gluons, etc) are not too important.

Next, define mixing coefficients (in our case, they are real):

$$v^a{}_n \equiv < \Phi^a \mid n >=< n \mid \Phi^a > \quad (4)$$

Neglecting the exponential contamination of order $e^{-E_{M+1}T}$, we see that the multistate coupling matrix can be fit to an expression of the form

$$C^{ab}(T) = \sum_{n=1}^{M} v^a{}_n v^b{}_n e^{-E_nT} \quad (5)$$

Of course, we cannot hope to extract $M$ independent time-dependencies with $N < M$ smearing wavefunctions, so only $N \geq M$ will be considered. Typically we shall extract the maximum information from the lattice data by picking $N = M + 1$ (the extra operator being the local current needed for the extraction of $f_B$).

The fit is performed by a chi-square minimization of

$$\chi \equiv \sum_{T=T_<}^{T_>} \frac{\mid C^{ab}(T) - \sum_{n=1}^{M} v^a{}_n v^b{}_n e^{-E_nT} \mid^2}{\sigma^{ab}(T)^2} \quad (6)$$

with respect to the fitting parameters $\{v^a{}_n, E_n\}$, over a fitting range $T_< \leq T \leq T_>$ in Euclidean time. The fit is performed on an ensemble of $N_c$ jack-knife coupling matrices obtained by replacing each in turn of the $N_c$ coupling matrices by the average matrix and reaveraging. We have chosen $| \Phi^N > \equiv J_{\rm axial}(0) | 0 >$, so the parameters ${v^N}_n$ should be interpreted as *lattice* $f$-parameters for the ground and excited meson states, $E_n$ as the corresponding masses, and ${v^a}_n$ as mixing coefficients indicating the degree of overlap of the exact meson states with our RQM smeared states $| \Phi^a >$. Note that this fitting procedure automatically gives the lattice $f$-parameters without the need to divide by the square-root of the smeared-smeared correlator as in the usual approach.

Once the overlaps $< \Phi^a | n >$ have been estimated by a best fit of $C^{ab}(T)$, a smearing operator can be constructed which is guaranteed to contain *at most one* of the first $M$ exact meson states, thereby removing any other exponential time-dependence to the $e^{-E_{M+1}T}$ level. Specifically, if $\epsilon_{a_1 a_2 .. a_M}$ is the totally antisymmetric symbol in $M$ dimensions, the smeared state

$$| \hat{\Phi}^A > \equiv \epsilon_{a_1 a_2 .. a_M} \prod_{i \neq A} {v^{a_i}}_i | \Phi^{a_A} > \quad (7)$$

is guaranteed (to the extent that we have accurately extracted the mixing coefficients ${v^a}_n$) to contain only the exact meson state $| A >$, together with contaminations from the $M + 1$'th excited state and higher. An effective mass plot of the usual kind can then be obtained for the A'th state by displaying

$$m^A_{\rm eff}(T - \frac{1}{2}) \equiv \ln \frac{C^A(T-1)}{C^A(T)} \quad (8)$$

where

$$C^A(T) = < \Phi^{(\rm loc)}, T | \hat{\Phi}^{(A)}, 0 > \quad (9)$$

The smeared-smeared correlator is defined in a similar fashion. We depict smeared-local rather than local-smeared correlators since the former have less noise.

## 4. PERTURBATIVE CORRECTIONS AND UNITS CONVERSION

Several parameters must be determined before one can convert current-current lattice correlators into physical numbers. Various methods for determining these parameters exist, and the state of the art is still advancing. In this section, we present the choice of parameters used in the present work. The information is summarized in the Table 1 below.

Table 1
Parameters for Lattice-to-Continuum Matching

| $\beta$ | $a^{-1}$(GeV) | $\kappa_c$ | $\delta\kappa^{-1}$ | $Z_A$ |
|---|---|---|---|---|
| 5.7 | 1.15(8) | 0.1691(1) | 0.253(4) | 0.679 |
| 5.9 | 1.78(9) | 0.15974(6) | 0.122(2) | 0.694 |
| 6.1 | 2.43(15) | 0.15495(4) | 0.080(1) | 0.702 |
| 6.3 | 3.05(15) | 0.15177(4) | 0.062(1) | 0.708 |

In Table 1, the five columns are the $SU(3)$ gauge coupling, $\beta$, the inverse lattice spacing in GeV as determined from the $1P$–$1S$ charmonium splitting[18] where available ($\beta = 5.7, 5.9, 6.1$) and by the $\rho$ mass for $\beta = 6.3$, the critical value of $\kappa$, the difference between $\kappa^{-1}$ for the strange quark and the critical value of $\kappa^{-1}$, and finally the value of $Z_A$, the perturbatively calculated renormalization factor which multiplies the lattice result. $Z_A$ is often taken to be 0.8, so the values for $Z_A$ given above, which represent as much as a 15% reduction from this merit further explanation.

There are two perturbative calculations which go into the determination of $Z_A$. They are a matching of the lattice regularized current to a continuum heavy quark effective theory current, and the running and matching of the latter current to a current in a theory with the full action for the heavy quark. The first of these two calculations has been tadpole-improved by Hernandez and Hill[2] following the prescription of Lepage and Mackenzie[11]. See also the discussion of this prescription to the heavy quark effective theory by Claude Bernard in these proceedings[3]. It is this tadpole-improvement that is responsible for the large change. The second part of the calculation has also been redone by the present authors to use the two-loop results of Ji and Musolf and Broadhurst and Grozin[12], but since this is not the origin of the largest part of the change in the size of $Z_A$ we will defer discussion of the two-loop calculation to another setting[13].

The tadpole improvement of the first of these two calculations using the prescription of Lepage and Mackenzie[11] has two major ingredients:
(1) A new lattice coupling $\alpha_V$ is defined. At a given scale, it is similar in size to $\alpha_{\bar{MS}}$, a well-behaved perturbative coupling, although it is defined directly from lattice calculations using the expectation value of the plaquette. The new coupling $\alpha_V$ is the same for all processes, but the scale at which it is evaluated is process-dependent. The scale has been determined for the process of interest here in reference [2]. The coupling $\alpha_V$ so determined is larger than $\alpha$ determined by $g^2 = 6/\beta$, and it is because of this that $Z_A$ differs from its tree-level value (unity) more than the often used value of 0.8.
(2) A reorganization of perturbation theory for the operator of interest, in this case, the transition of a heavy quark to a light quark caused by the weak axial vector current. The motivation for this reorganization [11] and the application of the prescription for heavy quarks on the lattice [3] may be found elsewhere. Here we simply give the final prescription we need.

Assuming one does the usual fitting procedure with current-current correlators (equation (21) in reference [15]), and that one is using the corresponding reduced value of the heavy quark wave function renormalization, there is no additional effect of tadpole improvement on heavy quark correlators. For an operator that involves both heavy and light (Wilson) quark fields, one must still take into account the effect of tadpole-improvement of the Wilson fermion action. The effect is that for each Wilson fermion in an operator, one should multiply the operator by the ratio

$$(8\kappa_c)^{1/2}_{\text{perturbative}}/(8\kappa_c)^{1/2}_{\text{non-perturbative}} \qquad (10)$$

The numerator should be evaluated to the order in perturbation theory as the other graphs contributing to $Z_A$ were computed. In the present case, this is one-loop. The one-loop values of $Z_A$ given above include this ratio. The non-perturbative values for the denominator that were used are those given by $\kappa_c$ in Table 1. To this order, no other corrections to the parameters coming from the lattice calculation need be applied.

## 5. EXTRACTION OF EFFECTIVE MASSES

To extract results for masses and decay constants we have used the set of gauge configurations and light quark propagators enumerated in Table 2. The light quark action we use is not $O(a)$ improved. In Table 2, the four columns are the gauge coupling, $\beta$, lattice size, the number of independent gauge field configurations, and the light quark $\kappa$ values calculated for each configuration. In this section, we illustrate the quality of results obtained using the spinless relativistic quark model wave functions by reproducing representative effective mass plots from $\beta = 6.1$ and $\beta = 5.7$.

Table 2
Configurations and Light Quark Parameters

| $\beta$ | lattice | confs | $\kappa$ |
|---|---|---|---|
| 5.7 | $12^3 \times 24$ | 100 | $.168, .165, .161$ |
| 5.9 | $12^3 \times 24$ | 100 | $.159$ |
| 5.9 | $16^3 \times 32$ | 100 | $.159, .158, .157,$ $.156, .154$ |
| 5.9 | $20^3 \times 40$ | 100 | $.159$ |
| 6.1 | $24^3 \times 48$ | 50 | $.1545, .154, .151$ |
| 6.3 | $32^3 \times 48$ | 50 | $.1515, .1513, .1500$ |

The hallmark of a pure, isolated ground state meson is an effective mass plot which is constant in time. In Figs. 1 and 2, we show our results for both the $SS$ and $SL$ local effective mass plots at $\kappa = 0.151$, $\beta = 6.1$. The errors bars plotted are statistical errors obtained from single elimination jackknife. In each effective mass plot, the fit interval is indicated by the range over which the fitted value of the mass is drawn.

The smeared-local effective mass reaches its asymptotic value around $T = 3$, while the smeared-smeared propagator is nearly asymptotic after $T = 2$. The results exhibit a single consistent plateau for both smeared-smeared and smeared-local propagators over a large range of $T$.

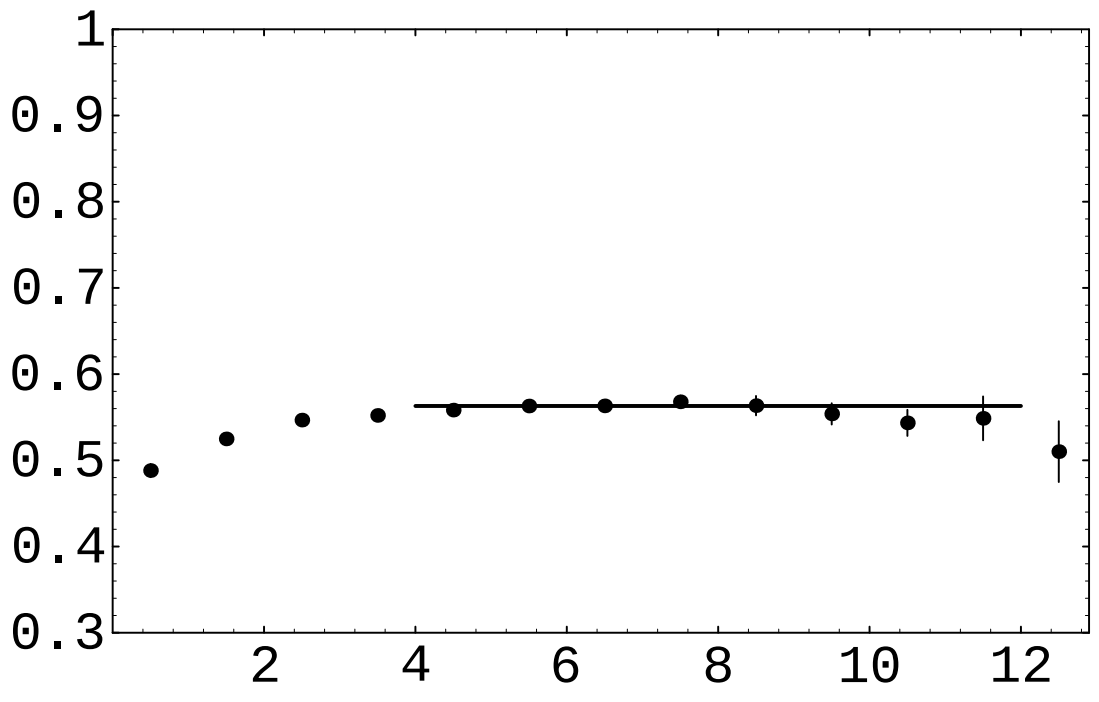

Figure 1. $\beta = 6.1$, $\kappa = 0.151$ Smeared-Local Effective Mass

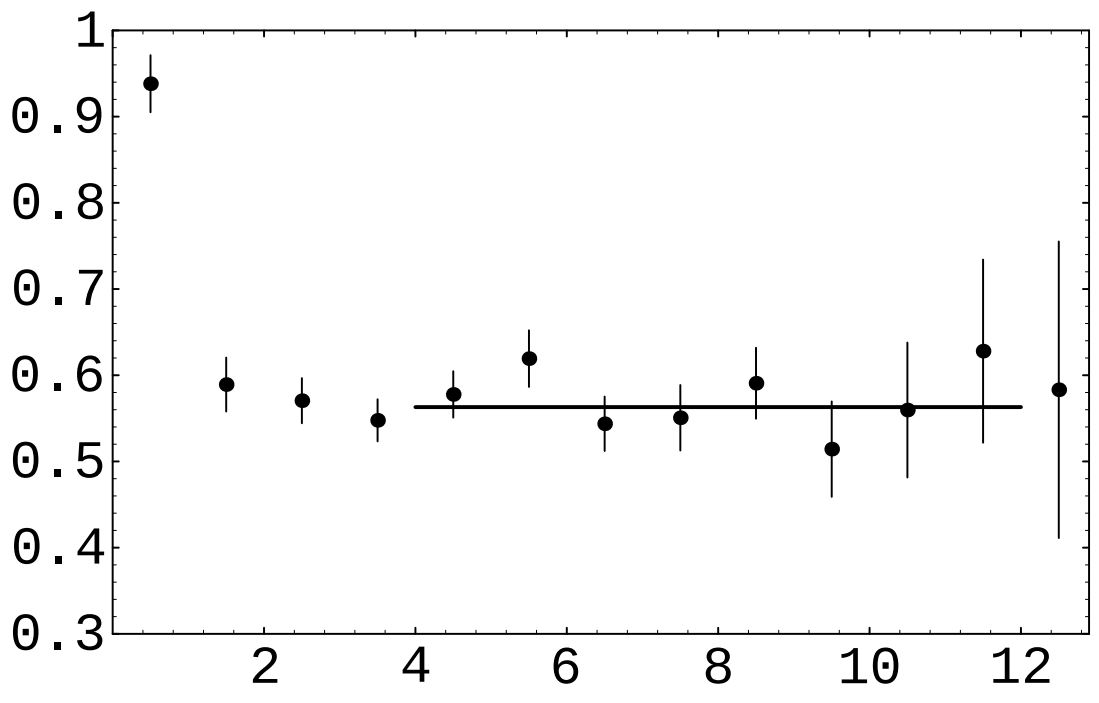

Figure 2. $\beta = 6.1, \kappa = 0.151$ Smeared-Smeared Effective Mass

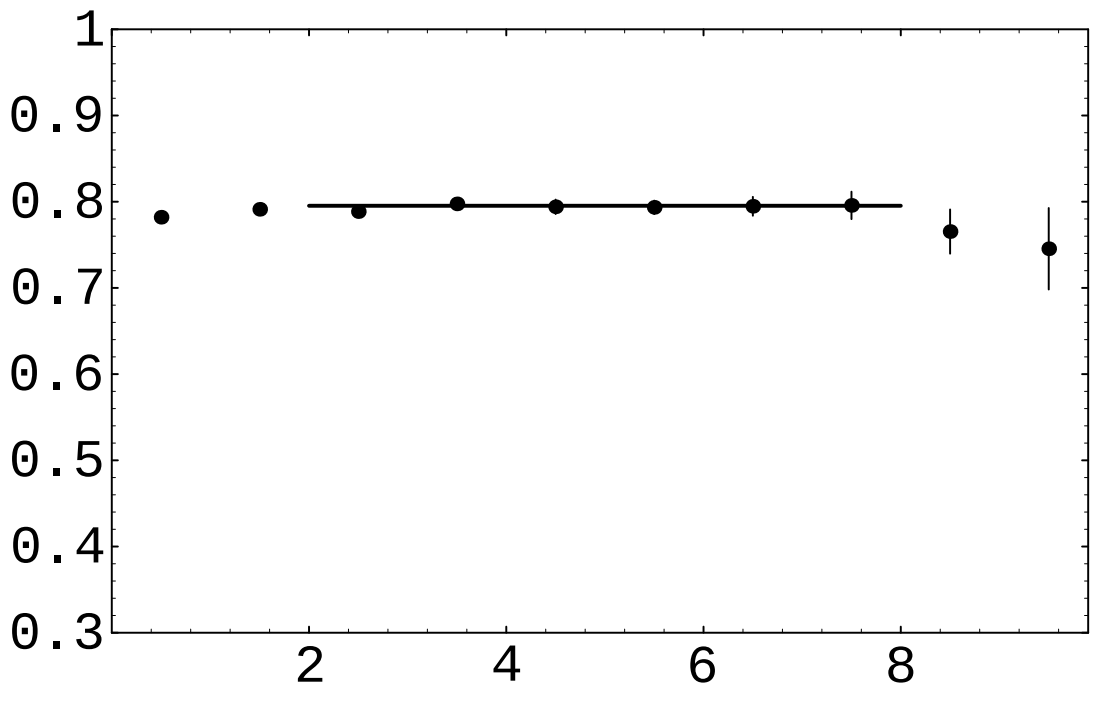

Figure 3. $\beta = 5.7, \kappa = 0.165$ Smeared-Local Effective Mass

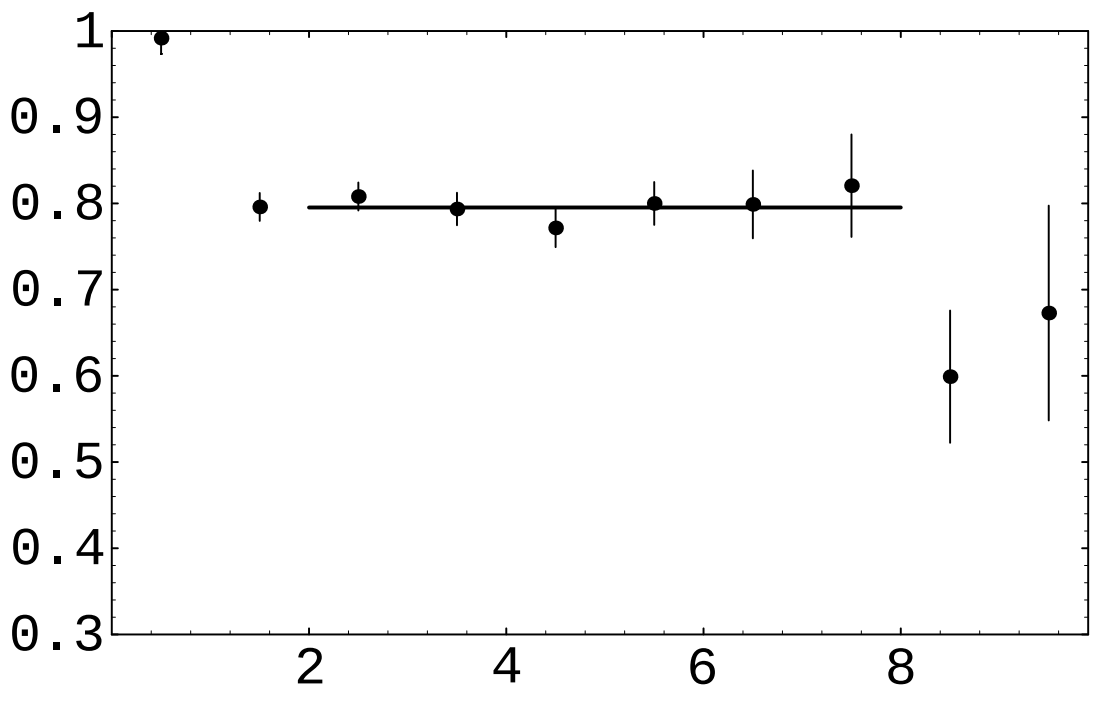

Figure 4. $\beta = 5.7, \kappa = 0.165$ Smeared-Smeared Effective Mass

In Figs. 3 and 4, we show our results for the $SS$ and $SL$ local effective mass plots at $\kappa = 0.165$, $\beta = 5.7$.

In general, the ranges were chosen to be approximately the same in physical units. The fit intervals 2–8, 3–10, 4–12, and 5–14 were used for the $\beta$ values 5.7, 5.9, 6.1, and 6.3 respectively. These mass plots convincingly demonstrate the effectiveness of our smearing method in isolating the ground state.

## 6. MASSES AND DECAY CONSTANTS

We are now ready to discuss our results for the mass and decay constant for our new $\beta = 5.7$, 5.9, 6.1, and 6.3 lattices. A detailed discussion of the $\beta = 5.9$ results (including the comparison to our previously reported results) will be delayed to the next section.

The results for the mass and decay constant as a function of $\kappa$ at $\beta = 5.7$ are presented in Figures 5 and 6. The values of $\kappa$ corresponding to the strange quark mass and $\kappa_c$ are denoted by vertical lines. The errors bars on the individual $\kappa$ values were determined using single elimination jackknife. Using these values and jackknife errors, a linear regression for the decay constants and masses at the various $\kappa$ values was performed. The values and errors of the slope and intercept of the fitted line can be found in Tables 3 and 4.

Similarly, the results for the mass and decay constant as a function of $\kappa$ at $\beta = 6.1$ are pre-sented in Figures 7 and 8, and the results for the mass and decay constant as a function of $\kappa$ at $\beta = 6.3$ are presented in Figures 9 and 10. As $a \to 0$, a clear downward trend in $f_B$ can be observed in the tabulated results. In Figure 11, one can see the downward trend (as the continuum limit is approached) present in the results in Table 3. The discussion of the systematic errors is deferred until the conclusions.

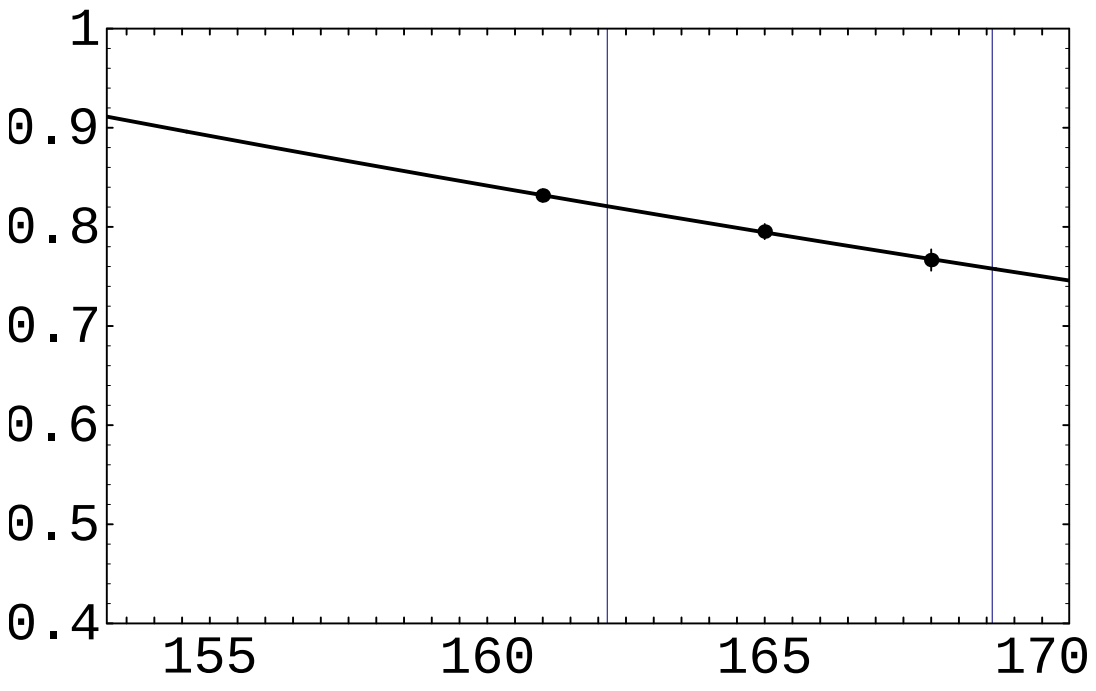

Figure 5. $\beta = 5.7$ Mass (Lattice Units) as a Function of $\kappa$.

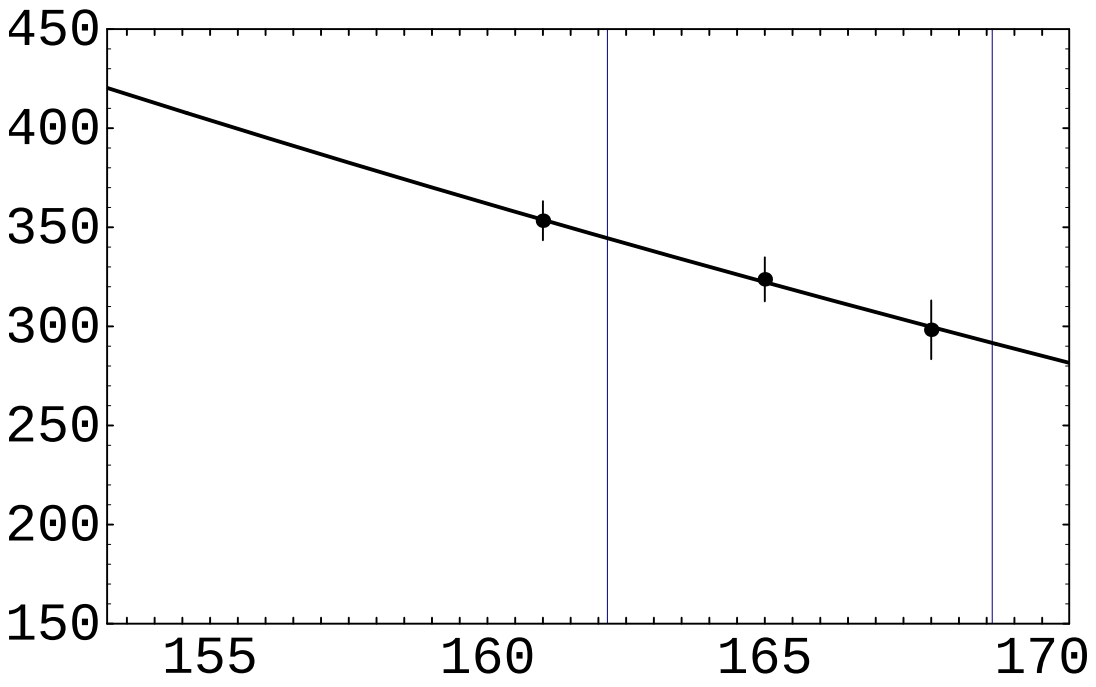

Figure 6. $\beta = 5.7$ Decay Constant (MeV) as a Function of $\kappa$.

## 7. $\beta = 5.9$ RESULTS

Preliminary results from a subset of the $\beta = 5.9$ gauge field configurations used here have previously been reported [10]. In this section, we report our results with twice as many gauge field configurations and the new multi-state analysis procedure, and compare with the aforementioned preliminary results.

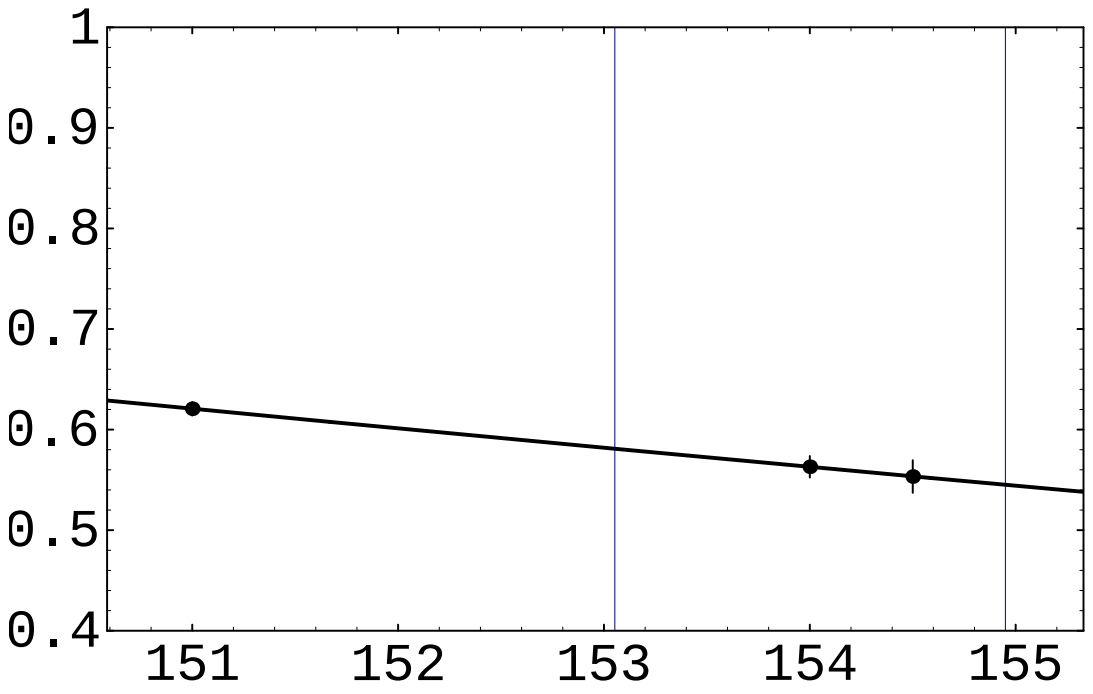

Figure 7. $\beta = 6.1$ Mass (Lattice Units) as a Function of $\kappa$.

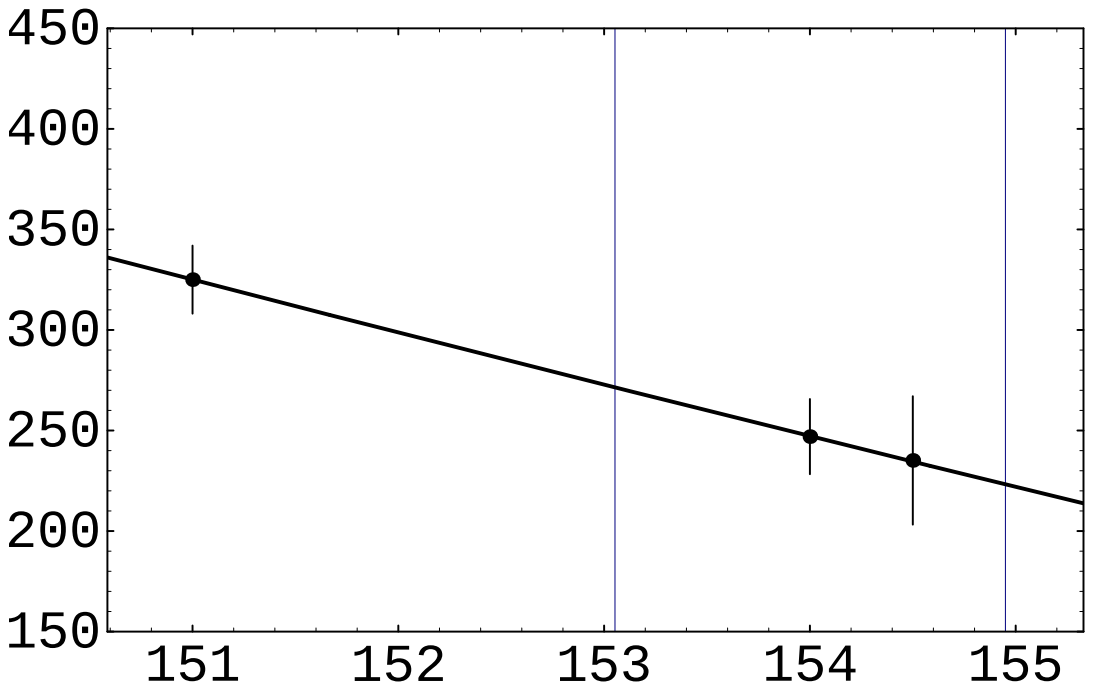

Figure 8. $\beta = 6.1$ Decay Constant (MeV) as a Function of $\kappa$.

To facilitate the comparison, we have replotted the preliminary results in Figure 12 after correcting for the reduction in $Z_A$ and a slight increase in $a^{-1}$. This changes the central value for the chiral extrapolation of $f_B$ from $319 \pm 11$(stat) MeV to $287 \pm 10$(stat) MeV. The results of the multi-state analysis are plotted in Figure 13. The result for the chiral extrapolation of $f_B$ of $268 \pm 14$(stat) MeV agrees to 1.1 standard deviations, but it is nevertheless worthwhile to comment on some substantial differences in the character of the results.

One notices several things about the compar-ison: (1) the central value of $f_B$ is reduced by somewhat more than one standard deviation, (2) the slope with respect to $\kappa$ is increased, (3) the final results have larger error bars, despite having twice the statistics, and (4) there is much less deviation from the linear fit in the final results.

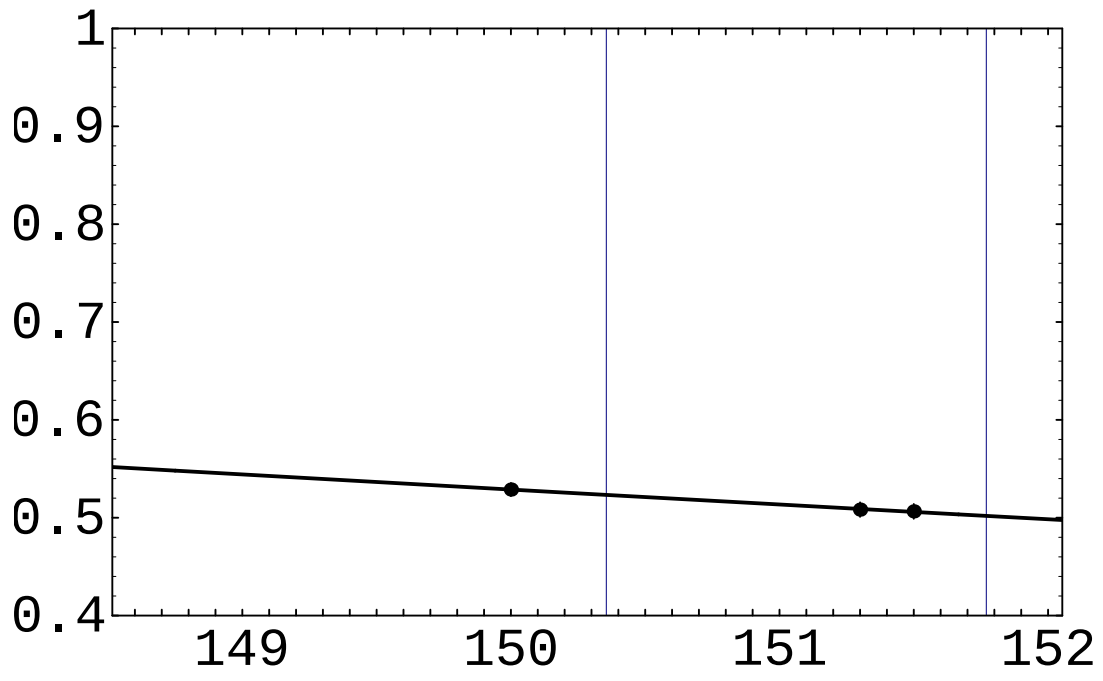

Figure 9. $\beta = 6.3$ Mass (Lattice Units) as a Function of $\kappa$.

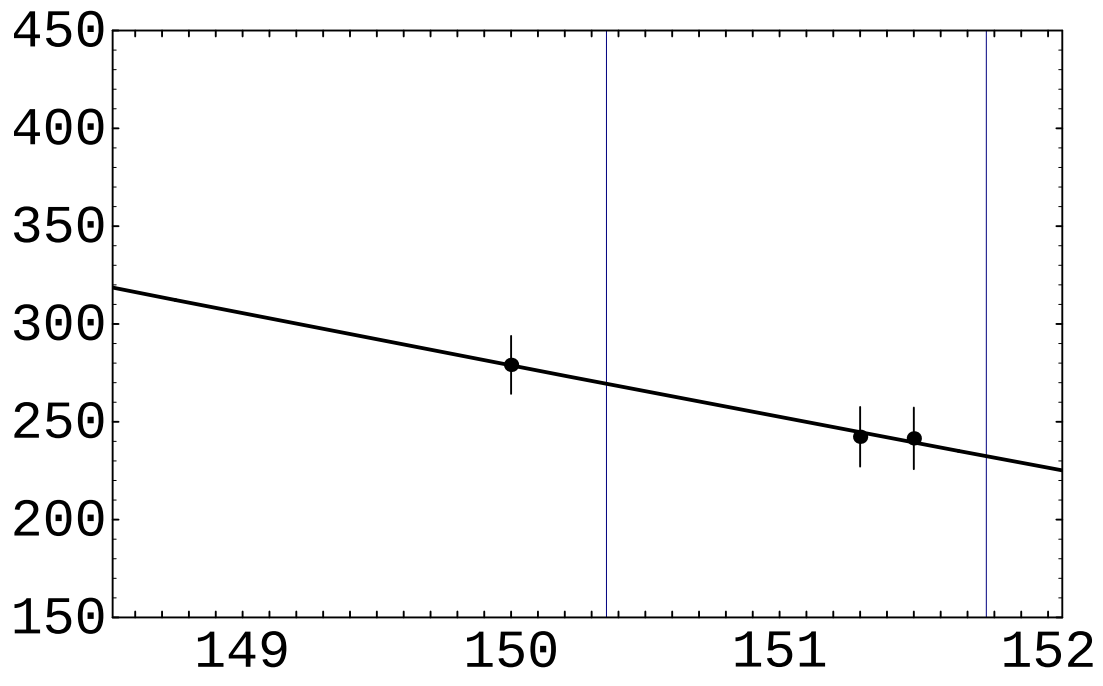

Figure 10. $\beta = 6.3$ Decay Constant (MeV) as a Function of $\kappa$.

The common origin of all four of these differences is that the new multi-state fitting procedure allows the admixture of excited states to vary from one jacknife subensemble to another. This converts a systematic error into a statistical error and thus the error bars are larger. Since the systematic error was potentially and in practice $\kappa$-dependent, this explains the deviation from a linear fit in the preliminary results, as well as the change in the central value of the slope. The new multi-state analysis eliminates one source of systematic error and replaces it with an increased statistical uncertainty in the results. The method is also more powerful in that it uses the full multi-state correlator at each time-slice and for each jackknife subensemble to determine the best fit values.

Table 3
Chirally Extrapolated Masses and Decay Constants

| $\beta$ | mass($a^{-1}$) | $f_B$(MeV) |
|---|---|---|
| 5.7 | $0.758 \pm 0.010$ | $292 \pm 14$ |
| 5.9 | $0.645 \pm 0.008$ | $268 \pm 14$ |
| 6.1 | $0.545 \pm 0.011$ | $223 \pm 21$ |
| 6.3 | $0.502 \pm 0.007$ | $232 \pm 14$ |

Table 4
Slopes with Respect to $\kappa^{-1}$ of Masses and Decay Constants

| $\beta$ | mass($a^{-1}$) | $f_B$(MeV) |
|---|---|---|
| 5.7 | $0.249 \pm 0.044$ | $209 \pm 65$ |
| 5.9 | $0.317 \pm 0.053$ | $367 \pm 94$ |
| 6.1 | $0.448 \pm 0.083$ | $603 \pm 173$ |
| 6.3 | $0.357 \pm 0.143$ | $597 \pm 298$ |

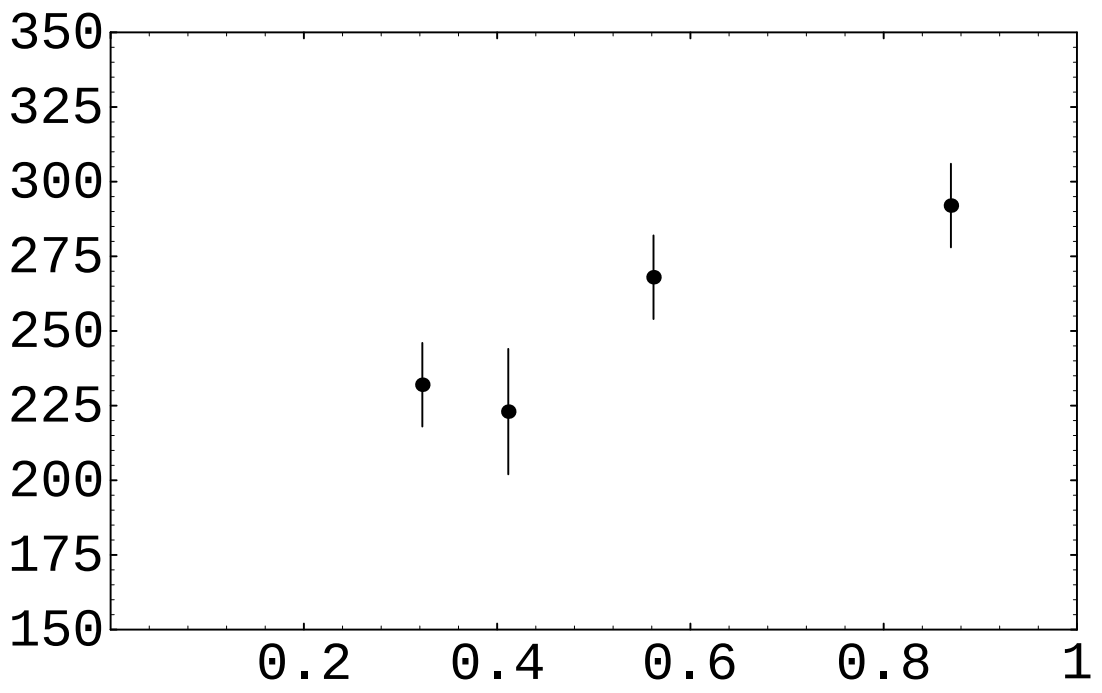

Figure 11. Decay constants (MeV) as a function of lattice spacing ($\text{GeV}^{-1}$)

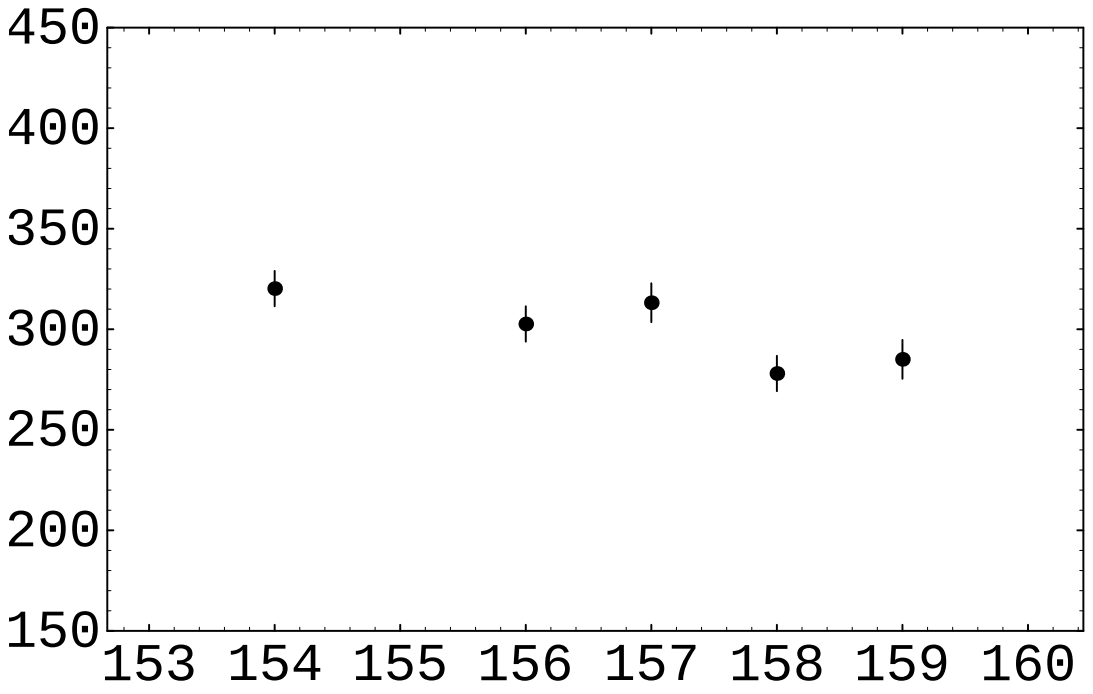

Figure 12. Previous $\beta = 5.9$ results for Decay Constant (MeV) vs. $\kappa$.

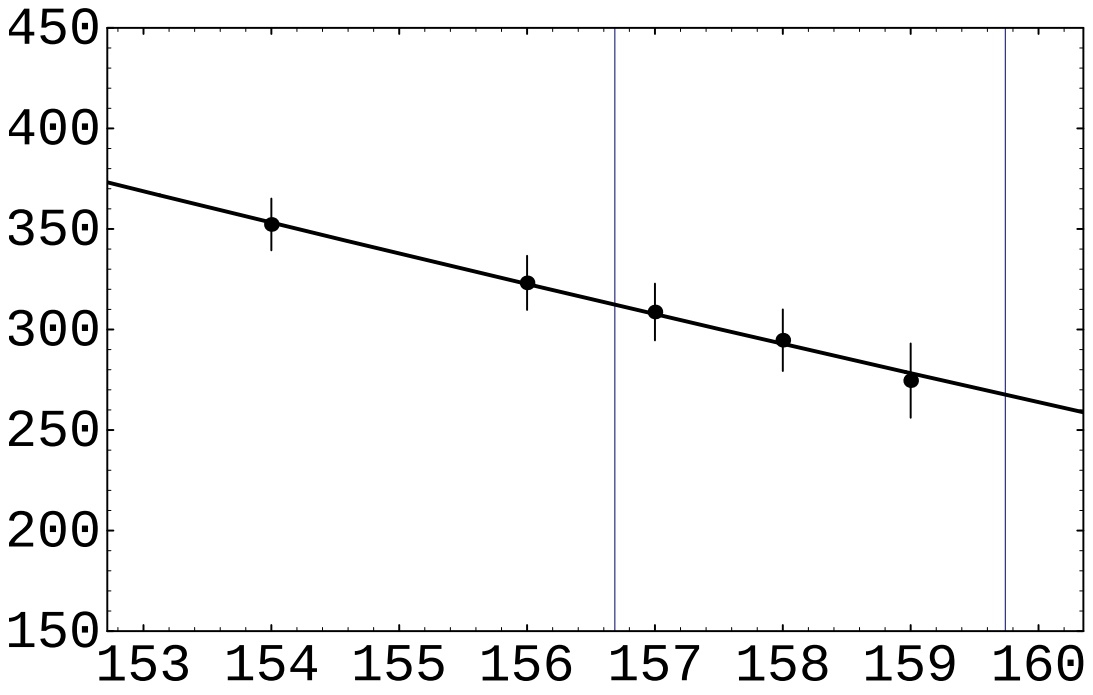

Figure 13. Present $\beta = 5.9$ results for Decay Constant (MeV) vs. $\kappa$.

## 8. CONCLUSIONS

We have presented results for the decay constant $f_B$ and for masses of low-lying heavy-light states in the static approximation. The analysis procedure introduces several improvements over previous smearing methods. First, the success of the RQM in reproducing the measured lattice wave functions is exploited by using the RQM to construct not only an accurate ground state wave function, but also a set of orthonormal excited state smearing functions. Second, the $\chi^2$ minimization procedure described in Section 3 makes full use of the information contained in the matrix of smeared-smeared and smeared-local correlators, including both ground state and excited state smearing functions at each end. Finally, our method provides much greater control over systematic errors from higher state contamination, because of the fact that the source smearing functions are tuned directly to the lattice wave functions, without regard to the behavior of the effective mass plots. The appearance of long plateaus in the SS and SL plots at the same value of effective mass is thus strong evidence that the systematic error from higher states has been largely eliminated. Additional evidence for this assertion has been obtained by comparing the results of a multistate calculation with $M = 2$ intermediate states to one with $M = 4$. The results for the ground state mass and decay constant show very little change between the 2-state and 4-state fits (always less than $\frac{1}{3}$ of the statistical error).

Table 5
Comparison of $B_s$ and $B_u$ Masses and Decay Constants

| $\beta$ | $m_{B_s} - m_{B_u}(MeV)$ | $f_{B_s}/f_{B_u}$ |
|---|---|---|
| 5.7 | 72(13) | 1.18(6) |
| 5.9 | 69(11) | 1.17(4) |
| 6.1 | 87(16) | 1.22(6) |
| 6.3 | 68(27) | 1.16(8) |

The results reported here only include statistical errors, the full analysis of the systematic errors will be reported elsewhere [13]. However all of our results are consistent with the conclusion that systematic errors from higher states are negligible compared to our statistical errors for the time interval used here. We have also studied the systematic errors associated with finite volume and extrapolation to $\kappa_c$ for the light quarks. As an example of the size of these errors, we estimate that for the $\beta = 6.3$ $f_B$ result in Table 3, the finite volume error is 4 Mev and the $\kappa_c$ error is 8 MeV. The $\kappa_c$ error represents our worse case and can be reduced by at least a factor of two when the smearing functions are properly optimized. Thus, we expect to be able to improve the accuracy of the present results by using larger ensembles. This would not be the case for previ-ously used smearing methods, which were dominated by systematic errors.

In addition to the overall scale uncertainty in the quenched approximation, there are two other sources of systematic uncertainty in our results for $f_B$. Use of the original Wilson action for the light quarks implies lattice spacing corrections in O(a) and the large one loop renormalization for the axial current suggests that the two loop correction may be sizable. In fact, one prominent feature of our results for $f_B$ at the four $\beta$-values studied is a rather strong dependence on the lattice spacing $a$ (c.f. Figure 11 and Table 3).

In Table 5 we present our preliminary results for the $B_s - B_u$ mass difference and for the ratio of decay constants for the $B_s$ and $B_u$. We see little $a$ dependence in either of these quantities. In particular, the systematic uncertainty due to the large perturbative renormalization of the axial current cancels for the ratio of decay constants. For both these physical quantities the extrapolation to the continuum limit ($a = 0$) is unproblematic.

JMF thanks the Nuffield Foundation for support under the scheme of Awards for Newly Appointed Science Lecturers. with the U.S. Department of Energy. BRH was supported in part by the Department of Energy under Grant No. DE–FG03–91ER 40662, Task C. HBT was supported in part by the Department of Energy under Grant No. DE-AS05-89ER 40518. AD was supported in part by the National Science Foundation under Grant No. PHY-90-24764. This work was performed using the ACPMAPS computer at the Fermi National Accelerator Laboratory, which is operated by Universities Research Association, Inc., under contract DE-AC02-76CHO3000.